\documentclass[fleqn,usenatbib]{rasti}

\usepackage{newtxtext,newtxmath}

\usepackage[T1]{fontenc}

\DeclareRobustCommand{\VAN}[3]{#2}
\let\VANthebibliography\thebibliography
\def\thebibliography{\DeclareRobustCommand{\VAN}[3]{##3}\VANthebibliography}

\usepackage{graphicx}	
\usepackage{amsmath}	
\usepackage{siunitx}
\usepackage{gensymb}	
\usepackage{color}

\usepackage{algorithm}
\usepackage{algpseudocodex}

\newcommand*{\Dex}{\emph{DexRT}}

\newcommand*{\Ha}{H$\alpha$}
\newcommand*{\Lya}{Ly$\,\alpha$}
\newcommand*{\Lyb}{Ly$\,\beta$}

\newcommand*{\LogEmis}{\ln{\left( \eta \Delta s\right)}}
\newcommand*{\LogOpac}{\ln{\left( \chi \Delta s \right)}}

\newcommand*{\Deleted}[1]{}
\newcommand*{\Replaced}[2]{#2}
\newcommand*{\Added}[1]{#1}

\newcommand*{\dd}{\mathop{}\!\mathrm{d}}

\title[A Non-LTE Ray Acceleration Structure]{A Simple Ray Acceleration Structure for Non-LTE Radiative Transfer}

\author[C. M. Osborne]{
C. M. J. Osborne$^{1}$\thanks{E-mail: Christopher.Osborne@glasgow.ac.uk}
\\
$^{1}$SUPA School of Physics and Astronomy, University of Glasgow, Glasgow, G12 8QQ
}

\date{Accepted XXX. Received YYY; in original form ZZZ}

\pubyear{2025}

\begin{document}
\label{firstpage}
\pagerange{\pageref{firstpage}--\pageref{lastpage}}
\maketitle

\begin{abstract}
We present a novel ray acceleration structure for \Deleted{non-local thermodynamic equilibrium (non-LTE)} radiative transfer\Added{ outside of local thermodynamic equilibrium (non-LTE)}, leveraging techniques from computer graphics to improve computational efficiency. By applying mipmapping (local recursive spatial averaging) and sparse voxel grids, we exploit spatial coherence and sparsity in astrophysical models to accelerate the formal solution of the radiative transfer equation. We introduce a variance-limited mipmapping (VLM) scheme with tunable error control, and extend it to handle anisotropic emission via two methods: velocity interpolation, and so-called "Core and Voigt". Our approach integrates a hierarchical digital differential analyzer (HDDA) for efficient ray traversal, which, combined with the mipmapping scheme achieves an order of magnitude speedup with less than $0.5\,\%$ error in the 99.9th percentile of the level populations. These methods are implemented in the DexRT code and demonstrate significant performance gains in realistic solar atmospheric models.
\end{abstract}

\begin{keywords}
Algorithms -- Radiative Transfer -- Spectral Lines
\end{keywords}



\section{Introduction}

As most astrophysical plasmas can only be investigated by observing their emitted radiation, radiative transfer is a key tool for modern astrophysics.
In particular, material outside of local thermodynamic equilibrium (so-called non-LTE), where the distribution of populations across atomic energy levels in the plasma is strongly influenced by the radiation field, remains a significant computational problem for realistic models, with potentially global coupling throughout the computational domain.
To determine the atomic energy distribution via the radiative transition rates, it is necessary to solve for the spectrum at multiple wavelengths per spectral line and continuum multiple times at each point in the volume, in an iterative manner.
\Added{In the time-dependent case, the radiation field itself is a 7-dimensional distribution, which varies over space, time, direction, and wavelength.}
Accelerating the calculation of the radiation field throughout \Replaced{the volume}{a domain} will provide significant gains for these iterative processes, as this so-called formal solution is the dominant cost of each iteration.
\Deleted{See \citet{Hubeny2014} for a thorough review of the techniques employed in the field of solar and stellar atmospheric radiative modelling.}
\Added{Whilst in this work we consider non-LTE radiative transfer, the techniques presented are general and could be applied to other problems, such as general scattering and neutron transport.}

\Replaced{Explicitly, the problem we are primarily interested in is referred to in computer graphics as volumetric rendering with participating media where the emissive and absorptive properties of the medium are known, potentially at fixed sample points.
Ignoring for now the effects of in-scattering (light originating from a different location and scattered along the direction of our observation) as this can be treated iteratively if the radiation field and scattering phase function are known, the volume can then be visualised by solving the radiative transfer equation along different rays \citep[reviewed in][along with stochastic methods for treating in-scattering]{pharr_physically_2023}, however for complex scenes this can be highly costly.
The monochromatic radiative transfer equation thus appears as
}{%
Explicitly, at each wavelength in the quadrature used for integrating the radiative rates, we solve the radiative transfer equation (RTE)
}
\begin{equation}
    \hat{\omega}\cdot \nabla I_{\lambda,\hat{\omega}} = \eta_{\lambda,\hat{\omega}} - \chi_{\lambda,\hat{\omega}} I_{\lambda,\hat{\omega}} \label{Eq:Rte},
\end{equation}
where $I_{\lambda,\hat{\omega}}$ is the specific intensity, $\eta_{\lambda,\hat{\omega}}$ and $\chi_{\lambda,\hat{\omega}}$ are the monochromatic emissivity and opacity at wavelength $\lambda$ in the medium from a particular point in a direction $\hat{\omega}$, respectively.
\Added{Along a ray with direction $\hat{\omega}$, this is then a first-order differential equation written as
\begin{equation}
	\frac{\dd{}I_{\lambda,\hat{\omega}}}{\dd{}s} = \eta_{\lambda,\hat{\omega}} - \chi_{\lambda,\hat{\omega}} I_{\lambda,\hat{\omega}} \label{Eq:RteRay},
\end{equation}
where $s$ represents the spatial coordinate along the ray.
The RTE is typically solved in integral form
\begin{equation}
	\begin{split}
		I_{\lambda,\hat{\omega}}(s_1) = &I_{\lambda,\hat{\omega}}(s_0) \exp{\left( -\int_{s_0}^{s_1} \chi_{\lambda, \hat{\omega}}(s) \dd{}s \right)} \\
		&+ \int_{s_0}^{s_1} S_{\lambda,\hat{\omega}}(s^\prime) \exp{\left(-\int_s^{s_1} \chi_{\lambda, \hat{\omega}}(s^\prime) \dd{}s^\prime \right)} \dd{}s
	\end{split}
	\label{Eq:RteIntegral}
\end{equation}
by assuming some functional variation of the opacity $\chi$ and source function $S$ (the ratio of emissivity to opacity) and integrating this analytically between $s_0$ and $s_1$.
}

When solving for the radiative rates\Added{ at a sample point}, it is necessary to \Replaced{compute the radiation in multiple directions at each point in the volume, typically using a fixed angular quadrature.
}{integrate the radiation over all directions.
This is typically performed using a discrete ordinate method with a fixed angular quadrature.}
The common procedures for this, which have remained essentially unchanged for the last three decades, are long and short characteristics.
In long characteristics, rays are traced from each sample point, along each direction of the quadrature, to the boundary of the grid \citep[e.g.][]{jones_formation_1973-1,jones_formation_1973,mihalas_two-dimensional_1978,de_vicente_long_2021}, whereas in short characteristics the radiation field is computed by sweeping across the domain for each direction in the quadrature, with each ray gaining upwind information via interpolation of the swept field and medium properties \citep{Kunasz1988}.
\Replaced{The short characteristics method is thus}{For a solution at the same sample points, the short characteristics method is} significantly more computationally efficient ($O(N^3)$ vs $O(N^4)$ for a three-dimensional domain of side $N$), but is also diffusive due to the interpolation.
\Replaced{Meanwhile}{On the other hand}, long characteristics is an embarrassingly parallel algorithm, whilst short characteristics presents complex ordering requirements \Added{in many-core environments due to both sweep ordering and domain decomposition} \citep[see][for a potential \Added{distributed memory} parallelisation scheme]{Stepan2013}.

Recently, \citet{osborne_radiance_2025} proposed a new method for computing the radiation field in a discretised atmosphere, termed radiance cascades.
This method uses multiple angular quadratures, trading off high spatial resolution and low angular resolution in the near-field, for low spatial resolution and high angular resolution in the far-field.
The implementation presented in their work employs a simple long characteristics style solver, however, this is not inherent to the method, and was chosen for simplicity.

Fortunately, models of astrophysical plasmas typically present structure in thermodynamic parameters (e.g. temperature, density, pressure), with smooth variation over large regions, punctuated by strong variations or even discontinuities in other regions (e.g. around shock jumps).
In the current work, we draw on computer graphics techniques to exploit sparsity and spatial coherence in the atmospheric models to accelerate the radiative transfer formal solution, with minimal error using local recursive \Added{de}refinement (mipmaps), and demonstrate potential implementations of this in our radiance cascades code \Dex{} \citep{osborne_radiance_2025}.
To maintain accuracy in our calculations, we adopt a tunable error metric to determine how to sample mipmaps, and present several different schemes for handling \Replaced{the traversal problem}{anisotropic emissivity and opacity}.

\Replaced{We first describe the difficulties of averaging quantities, the basic concepts of mipmapping and sparse grids in Section~\ref{Sec:TraversalProblem}, and then describe our approach to controlling error in grids of isotropic and anisotropic emission parameters in Sections~\ref{Sec:Static} and \ref{Sec:HandlingTraversalProblem}, respectively.
In Section~\ref{Sec:RayAcceleration} our approach to leveraging this to improve the efficiency of ray traversal through the sparse grid is presented, and the combined performance improvement provided by this technique is demonstrated in Section~\ref{Sec:PerformanceComparison}.
We note that the approach presented is not unique to the radiance cascades solution, and discuss how it can be applied to some other radiative transfer schemes in Section~\ref{Sec:Discussion}.}{In the remainder of the Introduction, we describe the difficulties of averaging atmospheric quantities  (Section~\ref{Sec:TraversalProblem}), and the basics of sparse adaptive grids for radiative transfer (Section~\ref{Sec:Sparsity}). Then in Section~\ref{Sec:Static} we describe our adaptive derefinement method and ray tracing acceleration scheme for the case of a static model, before extending it to models with bulk flows and anisotropic emissivities and opacities in Section~\ref{Sec:Dynamic}. A comparison of the different methods implemented in \Dex{} is presented in Section~\ref{Sec:PerformanceComparison}. Finally, we discuss avenues for future improvement and extensions of this method in Section~\ref{Sec:Discussion}.}

\subsection{Traversal Problem}\label{Sec:TraversalProblem}

\begin{figure}
	\centering
	\includegraphics[width=0.8\columnwidth]{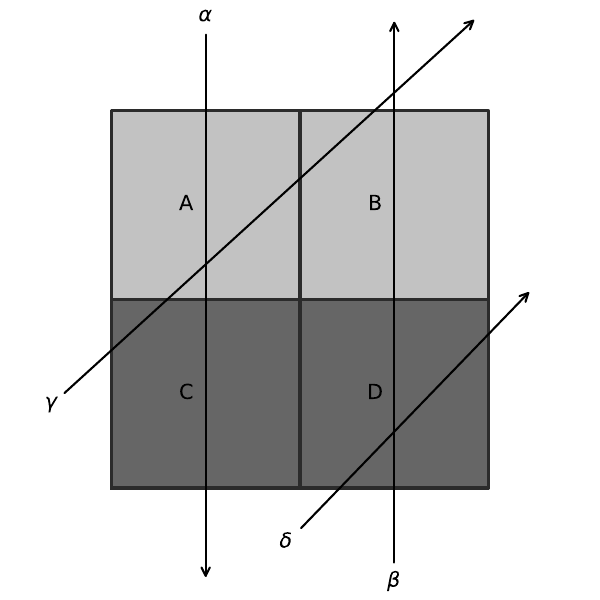}
	\caption{Illustration of the traversal problem in radiative transfer. If all voxels have non-trivial optical thickness with different emissivity and opacity (illustrated by A and B being \Replaced{red}{light grey}, and C and D being \Replaced{blue}{dark grey}) then the direction of traversal through the grid has a significant impact. Whilst propagating through equivalent voxels in a different order, the opposed rays $\alpha$ and $\beta$ see very different results. Similarly, rays $\gamma$ and $\delta$ which have the same direction but different origins see quite different results, due to only traversing a subset of the domain. \Added{A worked example is shown in Table~\ref{Tab:WorkedExample}.}}
	\label{Fig:TraversalProblem}
\end{figure}

\begin{table}
\centering
\begin{tabular}{lll}
\hline
Ray      & True & Average \\ \hline
$\alpha$ & 0.78 & 1.29    \\
$\beta$  & 2.00 & 1.29    \\
$\gamma$ & 2.65 & 1.25    \\
$\delta$ & 0.25 & 0.94    \\ \hline
\end{tabular}%
\caption{Intensity at heads of rays in Figure~\ref{Fig:TraversalProblem} for $\eta_A = \eta_B = 2$, $\eta_C = \eta_D = 0.4$, $\chi_A = \chi_B = 0.2$, $\chi_C = \chi_D = 1.4$, solving the RTE through A, B, C, and D separately, or for the path length through the average $\eta$ and $\chi$ in the four voxels.}
\label{Tab:WorkedExample}
\end{table}

Coherence and smooth varying thermodynamic parameters do not necessarily translate to smooth varying emissivity and opacity, especially in the case of non-LTE spectral lines\Added{ where emissivity and opacity are instead coupled to the global radiation field.
For instance, the emissivity and opacity of a spectral line forming in a slab with uniform thermodynamic parameters change rapidly in the outer layers as radiation starts to escape.}
In the following we consider that the distribution of emissivity and opacity throughout the model at a particular wavelength is known\Added{ (i.e. for an atomic transition, we assume that the atomic level populations governing emissivity and opacity in the transitions are known.)}.

For a model with a sharp spatial variation in emissivity or opacity a problem we term the \emph{traversal problem} appears.
This problem is illustrated in Figure~\ref{Fig:TraversalProblem}: it is clear that with non-trivial optical thickness the direction of traversal through a non-uniform grid of emissive parameters affects the intensity accumulated along the ray path.
\Added{Setting the emissivity and opacity in each voxel: $\eta_A = \eta_B = 2$, $\eta_C = \eta_D = 0.4$, $\chi_A = \chi_B = 0.2$, and $\chi_C = \chi_D = 1.4$ and taking a voxel side-length of 1, the intensity at the head each ray (assuming uniform parameters in each voxel and integrating \eqref{Eq:RteIntegral} accordingly) is shown in Table~\ref{Tab:WorkedExample}.
The ``Average'' column of this table also shows the results of integrating along these same paths, but taking the average emissivity and opacity across the four voxels for the complete path length.
For all four rays there are significant differences between the high-resolution solution and the one using the averaged parameters.}
\Deleted{By the same, if we were to average the emissive parameters across these cells, the solution of the radiative transfer equation along a ray is not in the general case the same as the average ray traversing the volume at full resolution, due to the non-linear nature of this integration.}
Nevertheless, it is clear that in the case of homogenous parameters \Replaced{these two formulations \emph{are}}{the averaged formulation is} equivalent, and the approximation does not become immediately invalid in the case of small variations in \Replaced{emission parameters}{emissivity and opacity} (as is common in many smooth regions of simulations).

\subsection{Sparsity \& Adaptive Resolution}\label{Sec:Sparsity}

To efficiently exploit a voxelised representation of a complex scene \Deleted{such as those typical in computer graphics or the solar prominence models \Deleted{presented }discussed by \citet{osborne_radiance_2025}, }sparse data structures can be powerful.
\Added{A good starting example is the classic octree which recursively subdivides parent voxels into 8 children on each successive level (equivalently a quadtree which subdivides into 4 children in 2D).
However, these structures are not particularly cache-friendly for modern hardware, and often lead to many redundant refinement levels that need to be traversed when storing and accessing high-resolution sparse data.}
\Replaced{Several of these have been proposed but typically revolve}{Computer graphics research has proposed several structures revolving} around the use of a sparse grid containing \Added{high-resolution, homogenous, potentially-nested} ``bricks'' of data e.g. \emph{brickmaps} \citep{christensen_irradiance_2004}, \emph{sparse voxel octrees} \citep{olick_next_2008}, \emph{gigavoxels} \citep{crassin_gigavoxels_2009,richermoz_gigavoxels_2024}, \emph{open and nano VDB} \citep{museth_vdb_2013,museth_nanovdb_2021}.
All of these structures now have high performance graphics processing unit (GPU) implementations, and are optimised by their use of ``bricks'' allowing for contiguous memory access in many cases.
\Replaced{Additionally, the VDB family of structures utilises a higher tree branching factor than the 8 used in a traditional octree to reduce the number of levels and amount of pointer chasing necessary to traverse the tree.}{The VDB\footnote{VDB is not an acronym.} family of grid structures consists of shallow trees (up to three levels), but much higher subdivision ratios than a classic octree.
In the default configuration the base level parent node contains $32^3$ children, the first sub-level $16^3$ per node, and finally $8^3$ leaves per parent on the finest level.
This leads to an effective potential $4096^3$ leaves per base level node.
It is a two-level VDB-like structure that our grid structure (presented in Section~\ref{Sec:RayTraceAcceleration}) is based on.}

Adaptive resolution schemes are \Replaced{not}{far from} unique to computer graphics, and are often employed to improve the efficiency of computational fluid dynamics \citep[e.g. the popular MPI-AMRVAC,][]{keppens_mpi-amrvac_2023} have also been applied to radiative transfer, both in the context of adaptive mesh refinement \citep[e.g][]{jessee_adaptive_1998,rijkhorst_hybrid_2006,juvela_loc_2020}, and also unstructured grids \citep[e.g.][]{DeCeuster2020a,udnaes_irregular_2023}.
For the common astrophysical case of steady-state solutions to the spectral line transfer problem, several key issues arise:
\begin{enumerate}
	\item The problem is potentially global. In the case of non-equilibrium radiative transfer, it is necessary to compute the \Replaced{radiance}{intensity} distribution throughout the volume to determine the radiative rates, and as such, every region can couple to every other region \Added{and can be difficult to computationally decompose}.
	\item Transparent regions (low opacity) can contain \Replaced{radiance}{intensity} distributions of arbitrary complexity, based on the incoming radiation at their boundaries. Two interacting regions separated by a transparent region should not be affected by the treatment in this region.
	\item Different \Replaced{spectral lines (of the same species)}{wavelengths (both core and wing of the same line, or different spectral lines)} can be produced in regions with different thermodynamic parameters (e.g. temperature/density), and thus require different mesh refinement.
\end{enumerate}

For example, the octree refinement and de-refinement method of \citet{juvela_loc_2020} can fail to correctly resolve (ii) as rays exiting refined regions are averaged into a single parent ray, and any internal spatial structure is lost upon encountering further refined regions.
The correctness of this approach must therefore be carefully validated on complex structures with strong variation in thermodynamic parameters and optical depth.
The unstructured schemes can violate (iii) if treating multiple spectral lines, as an optimal sampling of a thermodynamic structure may be different for different lines, for example, a very thick line such as \Lya{} \Replaced{may}{will typically} form over more compact, hotter, regions than \Ha{}.

\section{Adaptive Radiative Transfer in a Static Model}\label{Sec:Static}

\Added{Assuming the medium is static (without the presence of velocities) and the atomic level populations are known in each voxel of the full resolution grid, the emissivity and opacity of each voxel is isotropic and can be immediately computed for each wavelength considered.
Our goal is then to determine a wavelength-specific grid structure over which the radiative transfer equation can be evaluated more efficiently, but without introducing significant error.
}

\Added{We write the emissivity and opacity as the sum of a background term (subscript ``bg'', containing all continuum processes), and a set of lines $\mathcal{L}$, noting that in this case $\eta_{\lambda,\hat{\omega}}$ and $\chi_{\lambda,\hat{\omega}}$ in \eqref{Eq:Rte} are no longer a function of direction $\hat{\omega}$, i.e.
\begin{align}
	\eta_{\lambda,\hat{\omega}} &= \eta(\lambda) = \sum_{l \in \mathcal{L}} \eta_l (\lambda) + \eta_{\mathrm{bg}}(\lambda), \label{Eq:EtaTot} \\
	\chi_{\lambda,\hat{\omega}} &= \chi(\lambda) = \sum_{l \in \mathcal{L}} \chi_l (\lambda) + \chi_{\mathrm{bg}}(\lambda), \label{Eq:ChiTot}
\end{align}
where the expressions for the emissivity and opacity of line $l$, between atomic energy levels $i$ and $j$ ($j > i$) are given by
\begin{align}
	\eta_l(\lambda) &= \frac{hc}{4\pi\lambda} n_j A_{ji} \varphi_l(\lambda), \label{Eq:Eta} \\
	\chi_l(\lambda) &= \frac{hc}{4\pi\lambda} (n_i B_{ij} - n_j B_{ji}) \varphi_l(\lambda). \label{Eq:Chi}
\end{align}
Here $A_{ji}$, $B_{ij}$, and $B_{ji}$ are the Einstein coefficients for spontaneous emission, absorption, and stimulated emission, respectively, $\varphi_l$ is line absorption profile, and $n_i$ is the population of atomic energy level $i$.
These expressions are written with the assumption of complete frequency redistribution (CRD), i.e. assuming that the line emission profile is the same as its absorption profile, but our approach could be generalised to a treatment including partial frequency redistribution (PRD).
}

\Added{Given the decoupling of the atomic level populations from the model thermodynamic parameters (in the case of non-LTE), any adaptive radiative transfer scheme cannot be based purely on the local thermodynamic parameters. Further, given the non-linearity of the level populations on the formal solution of the radiative transfer equation, it appears ill-advised to base the adaptivity purely on the level populations.
Doing so also leads to a more complex solution as a number of overlapping lines can all contribute to the emissivity and opacity at a single wavelength, leading to a potentially large number of populations to consider. Instead we perform all adaptation on the emissivity and opacity, calculated at the maximum resolution of our grid structure, and then simplified for the solution of the RTE.}

\subsection{Mipmapping}

\begin{figure*}
	\centering
	\includegraphics[width=\textwidth]{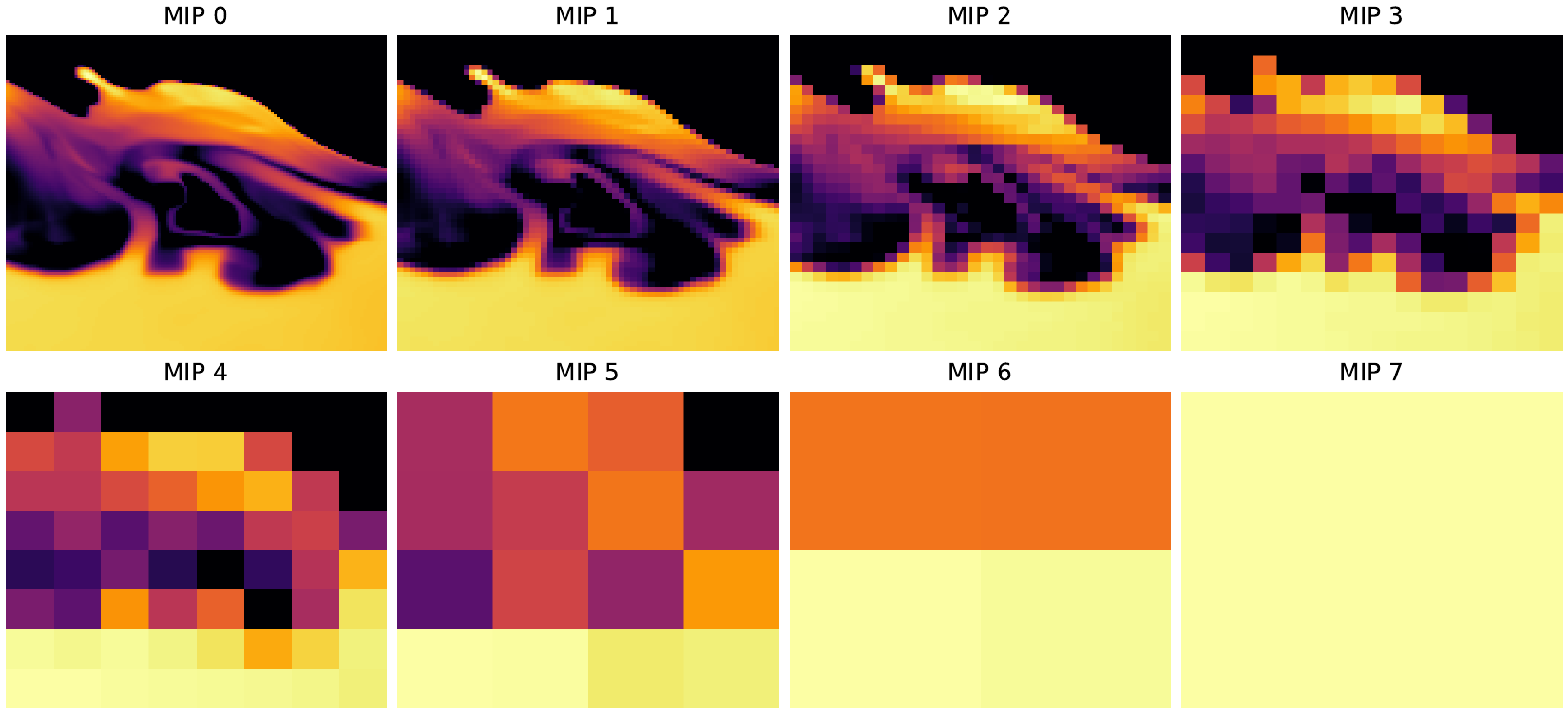}
	\caption{A demonstration of the process of successive spatial averaging in mipmapping \Added{applied to the opacity $\chi$ in the hydrogen \Lya{} line core in a model}. For each subsequent MIP level, the effective resolution is reduced by a factor of 2 on each spatial axis by applying a 2x2 box filter. Due to the decreasing size of the image with successive MIPs this set of images is known as a MIP pyramid\Added{ or MIP chain}.}
	\label{Fig:Mips}
\end{figure*}

\begin{figure*}
	\centering
	\includegraphics[width=\textwidth]{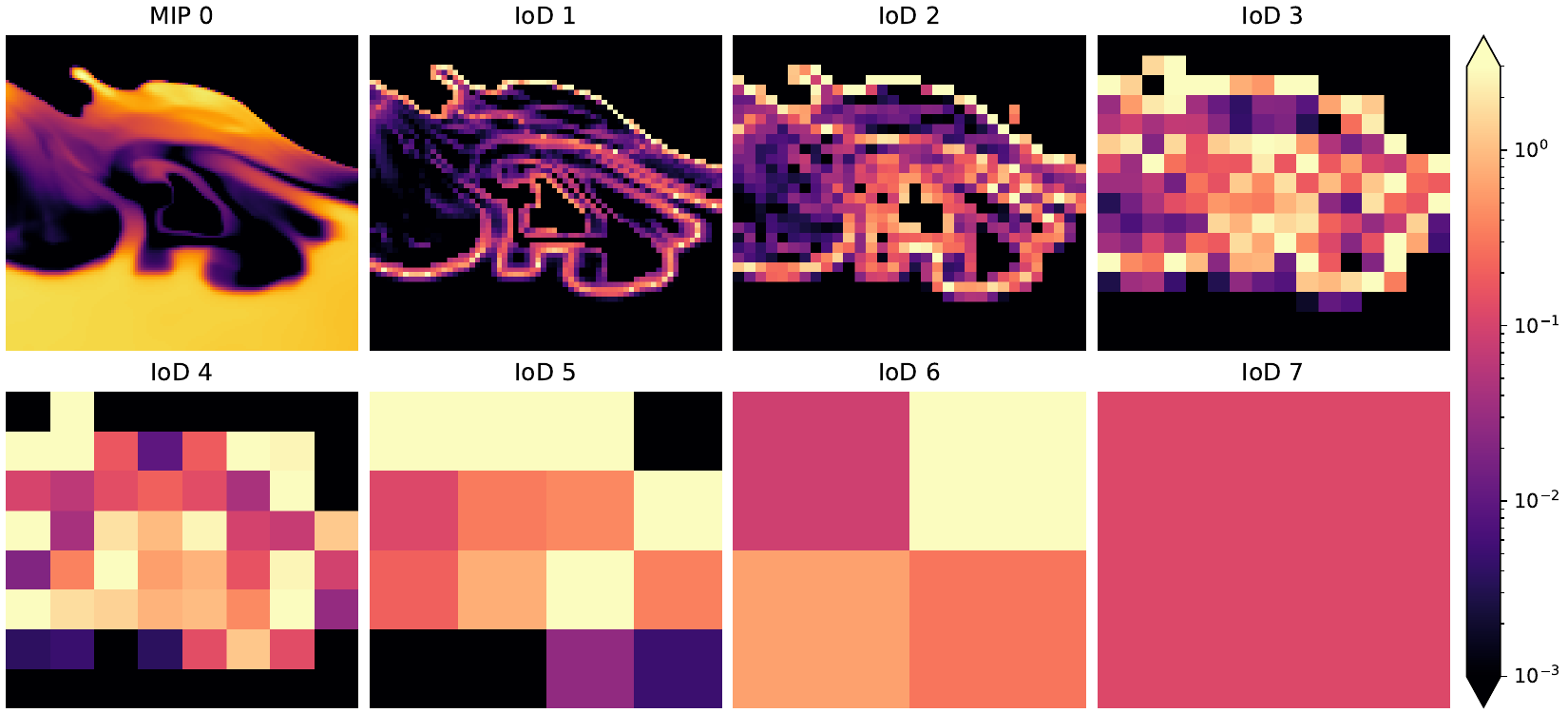}
	\caption{Values of the \Added{absolute value of the} index of dispersion (IoD -- variance of the \Replaced{pixels}{voxels} in the filter divided by the mean) for each MIP level of \Replaced{the red channel}{$\LogOpac{}$ given the opacity distribution in Figure~\ref{Fig:Mips}}. The higher this metric, the higher the variance in a pixel, and thus the more error will likely result from mipping this region to this level.}
	\label{Fig:IoD}
\end{figure*}

\Added{
Originating from the field of computer graphics, mipmapping\footnote{\emph{mip} coming from the Latin `multum in parvo' meaning `many in small'.} describes the process of constructing and sampling pyramids of reduced resolution data instead of just the full resolution dataset.
It is conventionally employed in the context of rasterising two-dimensional finite-resolution images from three-dimensional scenes where the scene objects are coloured based on two-dimensional images (textures) with a projected resolution that may differ dramatically from the resolution of the output image.
This pyramidal averaging of images (for mapping onto three-dimensional surfaces) was first described by \citet{williams_pyramidal_1983}.
Their construction of a pyramid of decreasing resolution for each texture increased computational efficiency by minimising the number of texture samples that need to be performed, but crucially reduced aliasing (such as Moiré patterns) due to mismatch between the sample frequency (screen pixels) and spatial data frequency (texture pixels).
Typically each layer of this pyramid is referred to as a MIP (with a full-resolution level of 0), and higher levels can be computed by applying different minification filters, however a simple averaging box filter is common.
An example set of MIPs generated from the hydrogen \Lya{} line core opacity ($\chi$) of a subregion of the model later employed for testing in Section~\ref{Sec:PerformanceComparison} is shown in Figure~\ref{Fig:Mips}.
We note that due to the successive halving of resolution, storing the entire MIP chain requires only $1.33\times$ the memory of MIP 0.
Applying this minification process equivalently in three-dimensions to produce \emph{volumetric MIPs} is even more effective, as the storage factor falls to only $1.14$.
For a MIP level $m$, each pixel or voxel has a side length $2^m$ of the full-resolution MIP 0.
}

\Deleted{Assuming for now that the emission and absorption parameters of the medium are known and isotropic in each voxel, we desire a simple method for determining if using the average properties of an $n\times n$ block of voxels will yield a good approximation to the solution of the radiative transfer equation through the full resolution of this block.
Following the principles of mipmapping shown in Figure~\ref{Fig:Mips}, we perform this calculation successively over blocks of $2\times 2\,(\times\,2)$ voxels in 2D (3D).
We have found that the index of dispersion}

\subsection{Variance Limited Mipmapping}\label{Sec:Vlm}

\Added{Whilst the emissivity and opacity can be directly mipmapped, given the non-linearity of the RTE, doing so can lead to significant errors, as discussed in Section~\ref{Sec:TraversalProblem}.
Instead we propose to determine, via a simple metric, the regions where the emissivity and opacity can be derefined without introducing significant error in the computed intensity.}

\Added{We find that the the absolute value of the index of dispersion (IoD)}
\begin{equation}
	\mathrm{IoD}[X] = \left| \frac{\mathrm{Var}[x]}{\mathrm{Mean}[X]} \right|
	\label{Eq:IoD}
\end{equation}
of $\LogEmis{}$ and $\LogOpac{}$, with $\Delta s$ the side length of a voxel, serves as a good indicator of quality of the mipmap approximation \Added{at a particular wavelength}.
Here, $X$ represents one of these parameters, $\mathrm{Var}[X]$ the variance, and $\mathrm{Mean}[X]$ the mean of the parameter.
\Deleted{This process is performed independently for different wavelengths in the model, allowing for estimation of the regions of high variation in different transitions.}
In Figure~\ref{Fig:IoD}, we show the value of the IoD \Added{for $\LogOpac{}$} over each of the successively averaged 4 \Replaced{pixels}{voxels} making up each MIP of \Replaced{the red channel of the pillars of creation image}{$\chi$ in Figure~\ref{Fig:Mips}}.

\begin{figure*}
	\centering
	\includegraphics[width=\textwidth]{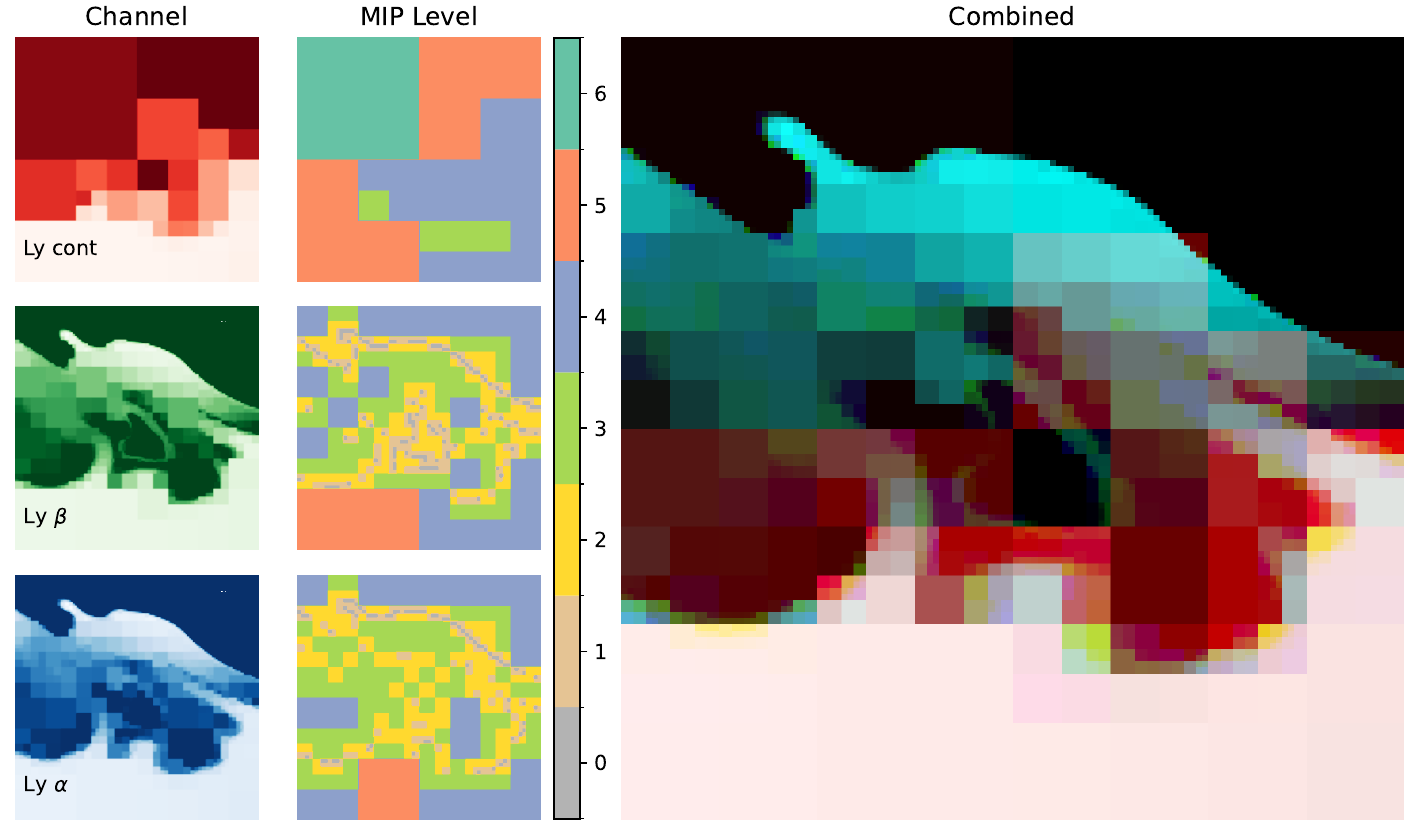}
	\caption{The result of applying our adaptive mipmapping technique independently to \Replaced{the three channels of the image}{three wavelengths of opacity of our model shown in Figure~\ref{Fig:Mips}}. Each \Replaced{channel}{wavelength} uses an IoD threshold of 1, and will not average further if a block reaches this threshold. From left to right we show each individually mipped \Replaced{channel}{wavelength}, the MIP level of each block, and the \Replaced{resultant image}{overlap of these mipmapped opacities (mapped into the red, green, and blue channels and scaled for visibility)}. The set of MIP levels shows that fine structure is present in different locations in the different \Replaced{channels: we see that there is overall more fine structure in the green channel}{wavelengths}. The fractions of \Replaced{pixels}{voxels} in each MIP level at each \Replaced{channel}{wavelength} is shown in Table.~\ref{Tab:MipFractions}}
	\label{Fig:AdaptiveMip}
\end{figure*}

\begin{table}
\resizebox{\columnwidth}{!}{%
\begin{tabular}{llllllll}
\hline
Channel & MIP 0 & MIP 1 & MIP 2 & MIP 3 & MIP 4 & MIP 5 & MIP 6 \\ \hline
Ly cont & 0.000 & 0.000 & 0.000 & 0.062 & 0.312 & 0.375 & 0.250 \\
\Lyb{}  & 0.030 & 0.092 & 0.167 & 0.242 & 0.344 & 0.125 & 0.000 \\
\Lya{}  & 0.018 & 0.063 & 0.208 & 0.336 & 0.312 & 0.062 & 0.000 \\ \hline
\end{tabular}%
}
\caption{Fraction of \Replaced{image pixels}{model voxels} in each MIP level for each \Replaced{channel}{wavelength} shown in Figure~\ref{Fig:AdaptiveMip}.}
\label{Tab:MipFractions}
\end{table}

We can construct \Replaced{a non-uniform image}{a problem-adapted grid} by successively averaging each spatial $2\times 2 \,(\times 2)$ block \Replaced{of each colour channel}{at a given wavelength}, and stopping if a particular \Deleted{threshold of} $\mathrm{IoD}$ \Added{threshold} is exceeded.
We refer to this process as Variance Limited Mipmapping (VLM).
\Deleted{The result of applying this process with an IoD threshold of 5 is shown in Figure~\ref{Fig:AdaptiveMip}.}
\Added{The result of treating the model shown in Figure~\ref{Fig:Mips} independently at 3 different wavelengths (Ly continuum head, \Lyb{}, and \Lya{}) and averaging blocks in each wavelength with $\mathrm{IoD}[\LogOpac{}] < 1$ is shown in Figure~\ref{Fig:AdaptiveMip}.}
\Deleted{We note that this is a very demanding image for this process due to all of the fine structure in the background starfield.
Nevertheless, all of the major structure of the image is retained, and less than 10\% of the pixels in any channel have remained at the \Replaced{base}{maximum} resolution (see Table~\ref{Tab:MipFractions}).}
\Added{It is interesting to note the varying refinement necessary to resolve the opacity distribution at these different wavelengths.
The right-hand panel of Figure~\ref{Fig:AdaptiveMip} shows the overlap between the structures at different wavelengths, highlighting regions that may be important at one wavelength, but not at another.
The breakdown of voxels between the different MIP levels is presented in Table~\ref{Tab:MipFractions}.
We note a significantly higher fraction of voxels are in MIP 0 and 1 to resolve \Lyb{} than for \Lya{}.
Under the metric adopted here, $<12\,\%$ of the voxels in any of these three wavelengths are located in the first two MIPs (which require 94\,\% of the storage in 2D).}

Empirically, we have found \Added{this} $\mathrm{IoD}$ threshold of $1$ for \Replaced{the log emissivity and opacity}{$\LogEmis{}$ and $\LogOpac{}$} to serve as a good trade-off between performance and accuracy\Deleted{, discussed further in Section~\ref{Sec:PerformanceComparison}}.
In practice, the IoD can become large even in region with no significant emission and absorption, so the $\mathrm{IoD}$ is only considered when $\chi \Delta s$ exceeds a threshold (typically 0.25), otherwise the region is always averaged.
\Added{At these lower values of $\chi \Delta s$ the exponential in the integral form of the RTE \eqref{Eq:RteIntegral} is reasonably approximated by the first-order Taylor expansion, and the expression becomes linear.}

\subsection{\Added{Tracing the Adapted Grid -- Ray Acceleration}}\label{Sec:RayTraceAcceleration}

\begin{figure}
	\centering
	\includegraphics[width=1.05\columnwidth]{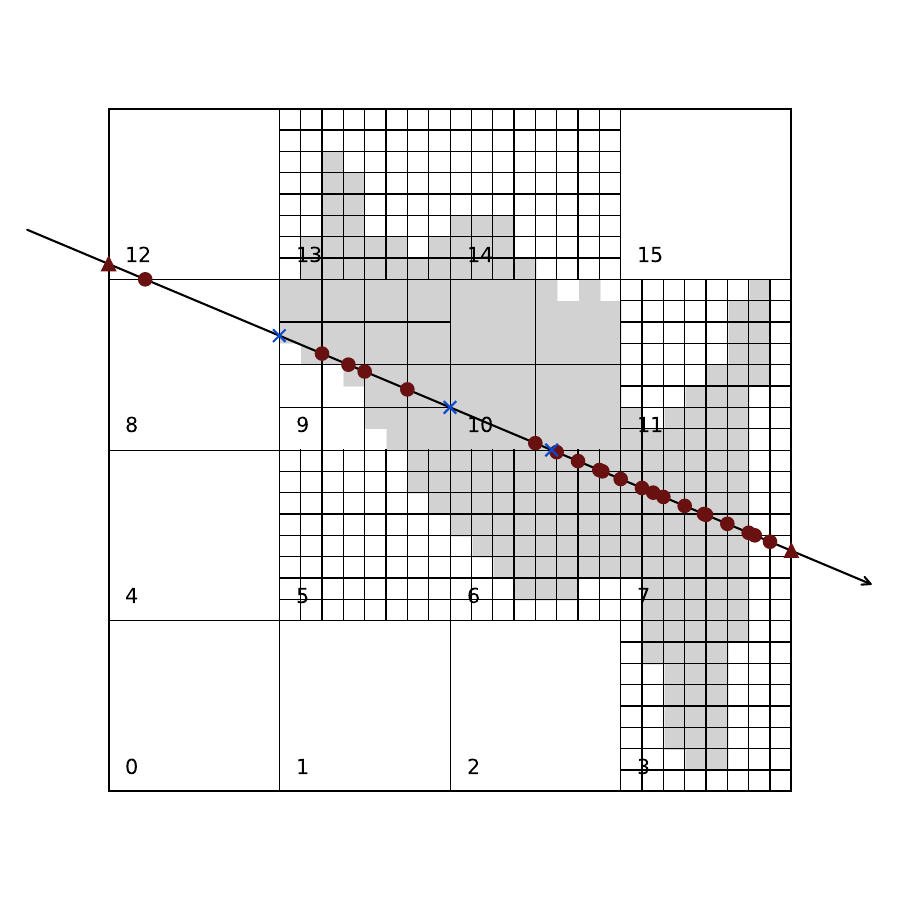}
	\caption{\Added{An example of our adaptive grid scheme and hierarchical digital differential analyser (HDDA) scheme, for a discretised model with with a full resolution of $32\times32$ (indicated in grey). The adaptive grid traced by the ray is determined as described in Section~\ref{Sec:Vlm}. Several ray intersection types are shown: with the model bounding box as red triangles, internally to the model as red circles, and on resolution change with blue crosses.}}
	\label{Fig:HDDA}
\end{figure}

\Added{In the previous section, we have outlined the construction of a problem-adapted grid for each wavelength of interest, however this structure needs to be efficient to raytrace for there to be any meaningful efficiency gain in the formal solution.
Our grid structure is primarily inspired by brickmaps \citep{christensen_irradiance_2004} and the VDB family \citep{museth_vdb_2013}, however many similar structures also exist.
The model is split into small cubic blocks -- we typically employ $16^2$ in 2D, and $8^3$ in 3D -- which may or may not be populated, and use a single refinement level per block.
A schematic 2D adapted grid is shown in Figure~\ref{Fig:HDDA}, but this process generalises directly to 3D.
The full resolution model (light-grey) is $32\times32$, and our grid employs a base resolution of $4\times4$ blocks where each of these blocks is a subgrid of varying resolution between 0 (empty), and $8\times8$ (to represent the full resolution of the model).
For example, blocks 0, 1, 2, 4, 8, 12, and 15 are empty, while block 9 is refined to $4\times4$ and block 10 to $2\times2$.
All other blocks are fully refined ($8\times8$).
}

\Added{Figure~\ref{Fig:HDDA} also shows an arbitrary ray crossing the domain, changing between various resolution blocks.
The red triangles indicate the intersections with the bounding box of the model (where model-specific boundary conditions will need to be applied), intersections where the ray changes resolution as blue crosses, and all other grid intersections as red circles.
Our ray traversal technique follows the Hierarchical Digital Differential Analyser (HDDA) of \citet{museth_hierarchical_2014} as an extension of the Digital Differential Analyser (DDA) presented in \citet[][henceforth AW87]{amanatides_fast_1987} to multi-resolution grids.
Our implementation assumes cubic voxels, but it can be equally adapted to uniform cuboidal voxels.
}

\Added{Rays are first clipped to the bounding box of the domain, and cast into characteristic form
\begin{equation}
	\vec{p}(t) = \vec{o} + t \vec{d},
	\label{Eq:Ray}
\end{equation}
where $\vec{o}$ is an origin point on the line defining the ray, $\vec{d}$ is its normalised direction vector, $t \in [t_1, t_2]$ with $t_1, t_2 \in \mathbb{R}$ and $t_2 > t_1$ such that $\vec{p}(t_1)$ is the ray start position, and the $\vec{p}(t_2)$ the ray end position.
The elegance of the AW87 method for uniform grids lies in its low floating-point arithmetic density and that the actual traversal is performed without evaluating \eqref{Eq:Ray}.
Our multi-resolution extension of this approach based on \citet{museth_hierarchical_2014}, minimises, but is not without evaluations of \eqref{Eq:Ray}.
It is defined over an index-space grid, where each maximum resolution voxel is of side-length 1.
An algorithmic overview of HDDA is given in Algorithm~\ref{Alg:HDDA}, where $d$ is the direction vector as per \eqref{Eq:Ray}, and \textsc{BitAnd} and \textsc{BitNot} refer to the bitwise and and not operations defined for integers. As such, $coord$ is an integer vector providing the $n$-dimensional index of the voxel containing $p(t)$, and the \textsc{NextPlaneHit} function returns $t$ for the intersection of the ray with the next downwind plane of constant coordinate $s$ (i.e. one of the bounding planes of the current voxel).
Whilst in a regime of constant voxel size, the ray proceeds by selecting the axis $s$ such that the plane of constant $s$ is the next voxel boundary intersected by the ray.
The ray then passes into the next voxel, having incremented its coordinate along axis $s$, and as the intersection with $next\_t[s]$ has now occured, this is incremented to the $t$ of the next ray intersection along this axis.
When the MIP level changes, the ray position within a voxel, and next intersection values are re-evaluated for the next MIP level, then tracing continues as before.
In the case of a single-resolution grid, $mip$ is always 0 and $vox\_size$ is always 1, and the traversal loop becomes extremely compact, essentially matching that of AW87, as the branch on line~\ref{AlgLine:ReinitMip} is never taken.
This is also true when using our scheme to march through a region of constant resolution -- this method will perform similarly to AW87 \emph{provided} the implementation of the \textsc{MipLevel} function is efficient.
This HDDA scheme provides the exact path length of the ray in each voxel, and does not resort to oversampling the domain with fixed-length steps as is performed in other ray-marching schemes \citep[e.g.][]{Osborne2019Thyr}.
}

\begin{algorithm}
\caption{Overview of the HDDA procedure.}
\label{Alg:HDDA}
\begin{algorithmic}[1]
\Function{FloorToStep}{$c, s$} \Comment{Floor vector $c$ to multiple of $s$}
	\For{each axis $a$ in $c$}
		\State $c[a] \gets \Call{BitAnd}{c[a], \Call{BitNot}{s - 1}}$
	\EndFor
	\State \Return $c$
\EndFunction
\\
\Function{MipLevel}{$c$} \Comment{Gets the mip level at coordinate $c$}
	\State $b \gets$ block containing $c$
	\State \Return \textbf{if} $b$ is empty \textbf{then} $-1$ \textbf{else} maximum MIP level of $b$
\EndFunction
\\
\Function{HDDA}{$t_1,t_2$}\Comment{Traverse $p(t)$ for $t\in[t_1, t_2]$}
	\State $t \gets t_1$ \Comment{Initialisation}
	\State $r \gets p(t)$
	\State $coord \gets \lfloor \mathrm{r} \rfloor$ \Comment{Per vector component}
	\State $mip \gets \Call{MipLevel}{coord}$
	\State $vox\_size \gets \Call{VoxelSize}{mip}$
	\State $coord \gets \Call{FloorToStep}{curr, vox\_size}$
	\For{each axis $a$ in $coord$}
		\State ${next\_t}[a] \gets \Call{NextPlaneHit}{a, vox\_size}$
		\State ${step}[a] \gets \mathop{\mathrm{sgn}}(d[a])$
		\State $\delta[a] \gets 1 / d[a]$
	\EndFor
	\\

	\While{$t < t_2$} \Comment{Traversal Loop}
		\State $s \gets \mathop{\mathrm{argmin}}(next\_t)$
		\State $t_h \gets \mathop{\mathrm{min}}(t_2, next\_t[s])$
		\\

		\If{$mip > -1$}
			\State Integrate RTE over $[t, t_h]$ in voxel $coord$
		\EndIf

		\\
		\State $t \gets t_h$
		\State $next\_t[s] \gets next\_t[s] + vox\_size * \delta[s]$
		\State $coord[s] \gets coord[s] + vox\_size * step[s]$
		\State $mip^\prime \gets \Call{MipLevel}{coord}$
		\\
		\If{$mip \neq mip^\prime$} \Comment{Reinitialise tracer for new MIP} \label{AlgLine:ReinitMip}
			\State $mip \gets mip^\prime$
			\State $r \gets p(t)$
			\State $coord \gets \lfloor \mathrm{r} \rfloor$
			\State $vox\_size \gets \Call{VoxelSize}{mip}$
			\State $coord \gets \Call{FloorToStep}{curr, vox\_size}$
			\For{each axis $a$ in $coord$}
				\State ${next\_t}[a] \gets \Call{NextPlaneHit}{a, vox\_size}$
			\EndFor
		\EndIf
	\EndWhile

\EndFunction
\end{algorithmic}
\end{algorithm}

\subsubsection{\Added{Implementation Details}}

\Added{In our implementation of HDDA we separate the storage of the block structure and the MIPs.
The block structure is most frequently referenced by the traversal code, and needs to be optimised to fit well in cache to allow for efficient skipping of empty space.
As such, the integer representation of the maximum permissible MIP level in each block (computed from the IoD thresholds) is stored using the minimum number of bits necessary to represent its complete range and packed into an array of 64-bit integers.
For simplicity, our implementation supports MIP levels where the maximum defined voxel size is the block size.
As an example, a block size of $16$ has a maximum MIP level of $4$, and $6$ potential states for its content: empty, or a MIP level in $[0, 4]$.
These $6$ states can be stored in $3$ bits, thus allowing the state of $21$ blocks to be packed into each 64-bit entry.
The block state of a $512^2$ model can then be stored in 49 64-bit entries (\SI{392}{\byte}), and equivalently 669 64-bit entries (\SI{5.4}{\kilo\byte}) for a $512^3$ model.
}

\Added{
In non-empty regions, it is necessary to access the MIP structure backing the blocks.
The full MIP structure for a quantity (e.g. $\eta$ or $\chi$) is allocated for every non-empty block, and for each MIP consists of a flat array of length $A B^D / 2^m$ where $A$ is the number of active blocks, $B$ is the block side length, $D$ is the dimensionality (2 or 3), and $m$ is the MIP level.
The data for each block is stored contiguously as a flattened Cartesian array with blocks laid out following a Morton z-order space-filling curve \citep{morton_computer_1966} to improve average cache access patterns along a ray.
When no blocks of a model are empty, the lookup from block coordinate to MIP storage location can be computed by computing its Morton code.
This is not possible in the case of arbitrary sparsity, and an offset lookup array is computed during MIP construction and VLM evaluation which stores the starting index in the MIP array associated with each block.
}

\Added{To reduce redundant calculations, we utilise lightweight objects to provide the index of a voxel in a MIP from a MIP level and coordinate.
These objects cache the block, and avoid recomputing the starting index of a block during repeated access to the same block.
}

\Added{The HDDA algorithm is implemented very similarly to pseudocode of Algorithm~\ref{Alg:HDDA}, however due to limited floating point precision during the repeated accumulation of terms into $next\_t$, it is possible for $p(t)$ to round slightly outside of $coord$ after traversing a large region of constant size voxels.
Two modifications have been made to the algorithm as written to make it robust to a large number of test cases in 2D and 3D:
\begin{itemize}
	\item Upon entering an empty region if $t$ has gathered sufficient error that $p(t)$ rounds back into the refined region, the traversal can enter an infinite loop getting stuck on this boundary due to the branch at line~\ref{AlgLine:ReinitMip}. To mitigate we increment $t$ by $10^{-2}$ of a voxel side length upon entering an empty region.
	\item $t$ must always increase when stepping along the ray. If $t_h <= t$ it is likely that there were two ray intersections very close to each other in $t$ (i.e. a ray passing through a voxel corner), as such $t_h$ is set to be very slightly larger than $t$.
\end{itemize}
}

\section{\Replaced{Handling the traversal problem}{Extension to Dynamic Models}}\label{Sec:Dynamic}

In realistic cases \Replaced{emission and absorption properties}{emissivity and opacity} are not isotropic due to velocity fields, and the traversal problem must be addressed in a more general way.
\Replaced{We will demonstrate two primary classes of solution to this problem: the former which cannot be applied directly here, and the latter building on our VLM approach described in Section~\ref{Sec:Vlm}.}{Here, we present an two extensions to the VLM approach described in Section~\ref{Sec:Vlm} for anisotropic quantities.
The first is a simpler, but more memory-intensive approach, while the latter is more complex and guarantees lower error.
We refer to both of these schemes as implementations of anisotropic variance limited mipmapping (AVLM).}

\subsection{Velocity Grids}\label{Sec:VelocityInterp}

\Replaced{In the complete frequency redistribution (CRD) case for spectral lines}{In the CRD regime}, the angular variation of \Replaced{emission parameters depends}{the emissivity and opacity depend} solely on the projected velocity.
As such the range of projected velocities \Replaced{in a}{for an arbitrary ray crossing a} voxel (which ranges between the positive and negative norm of the bulk velocity vector in that voxel) serves as a basis on which the \Replaced{anisotropy of emission and absorption parameters}{emissivity and opacity} can be computed and \Deleted{smoothly }interpolated.
\Replaced{Nevertheless, good}{A dense} velocity sampling is necessary to avoid large jumps \Replaced{in emission coefficients}{and inconsistencies in these terms, and this is the reason for the high memory cost of this approach}.

\Replaced{We compute a set of MIPs for velocities ranging between the minimum and maximum possible projection within each block, and interpolate based on the projected velocity along a ray during traversal.
To ensure consistency, a single MIP level is adopted per block, and is computed using the emission coefficients with all lines at rest for the wavelength.}{%
A set of emissivity and opacity MIPs is computed for a grid of $n$ velocities ranging between the minimum and maximum projections of velocity within each MIP voxel (i.e. either the maximum within a MIP 0 voxel, or over the set of voxels MIP 0 contained within a MIP $N$ voxel).
For a voxel containing material with bulk velocity $v$, the grid is thus uniform and linear between $-|v|$ and $|v|$.
Thus a full $n$ MIP chains for both emissivity and opacity at each velocity point in this grid are created and stored, along with a further $2$ MIP chains to store the parameters of the velocity grid in each voxel.}

\Added{The IoD metric that determines the MIP level used within a block is computed over the block using the values for the emissivity and opacity in the rest frame of each voxel (i.e. the centre of the velocity grid, with a velocity of 0) as it cannot be efficiently computed and sampled for every potential ray direction.
We recognise this may increase the error by using too high a MIP level in the wings of spectral lines where Doppler shifts cause excursions into the line core.
During ray traversal this grid of $n$ MIPs of emissivity and opacity is interpolated as a function of projected ray velocity to determine the approximate emissivity and opacity along the ray.}

The code uses a fixed number of \Replaced{wavelength}{linearly-spaced velocity} bins and warns if the spacing between bins is greater than two times (configurable) the thermal width of the spectral line with the closest rest wavelength\Added{ to the wavelength being considered}.
\Deleted{In general, this approach is fast, but quite memory intensive, due to the dense velocity sampling required.}

\subsection{``Core and Voigt'' method}\label{Sec:Cav}

\Replaced{We can additionally define an interpolation-free scheme for averaging and reconstructing the anisotropic emission properties without averaging or interpolating velocity-dependent quantities.
For a set of contributing spectral lines $\mathcal{L}$ at wavelength $\lambda$, with assumed isotropic background and continuum emissivity and opacity $\eta_{\mathrm{bg}}$ and $\chi_{\mathrm{bg}}$, we define the total emissivity and opacity
\begin{align}
	\eta_{\mathrm{tot}}(\lambda) &= \sum_{l \in \mathcal{L}} \eta_l (\lambda) + \eta_{\mathrm{bg}}(\lambda), \label{Eq:EtaCav} \\
	\chi_{\mathrm{tot}}(\lambda) &= \sum_{l \in \mathcal{L}} \chi_l (\lambda) + \chi_{\mathrm{bg}}(\lambda). \label{Eq:ChiCav}
\end{align}
The only directionally dependent component of the line emission coefficients is the line absorption profile, $\phi_l(\lambda, \hat{\omega})$, with direction $\hat{\omega} \in \mathbb{S}^2$.
Due to the CRD approximation, $\phi_l$ also serves as line emission profile, and considering only unpolarised non-relativistic radiative transfer, we can write
\begin{equation}
	\phi_{l}(\lambda, \hat{\omega}) = \varphi_l\left(\lambda + \cfrac{v \cdot \hat{\omega}}{c}\lambda_0\right),
\end{equation}
where $v$ is the plasma bulk velocity, and $\lambda_0$ is the rest wavelength of line $l$.
In practice, $\varphi_l$ is typically a Voigt profile with spatially varying damping factor for each line.}
{Modern many-core devices such as GPUs typically provide a very high ratio of compute throughput to memory-bandwidth.
As such, it can be advantageous to define an interpolation-free scheme for AVLM, building on the properties of the spectral lines.
Considering again the contributing set of spectral lines $\mathcal{L}$ for a dynamic atmosphere, with an isotropic background, we can now write the total emissivity and opacity from \eqref{Eq:EtaTot} and \eqref{Eq:ChiTot} as
\begin{align}
	\eta_{\lambda,\hat{\omega}} &= \sum_{l \in \mathcal{L}} \eta_l (\lambda, \hat{\omega}) + \eta_{\mathrm{bg}}(\lambda), \label{Eq:EtaTotDir} \\
	\chi_{\lambda,\hat{\omega}} &= \sum_{l \in \mathcal{L}} \chi_l (\lambda, \hat{\omega}) + \chi_{\mathrm{bg}}(\lambda). \label{Eq:ChiTotDir}
\end{align}
The line emissivity and opacity are now given by
\begin{align}
	\eta_l(\lambda, \hat{\omega}) &= \frac{hc}{4\pi\lambda} n_j A_{ji} \phi_l(\lambda, \hat{\omega}), \label{Eq:EtaDir} \\
	\chi_l(\lambda, \hat{\omega}) &= \frac{hc}{4\pi\lambda} (n_i B_{ij} - n_j B_{ji}) \phi_l(\lambda, \hat{\omega}), \label{Eq:ChiDir}
\end{align}
where $\phi_l$ is the line absorption profile in a moving medium at wavelength $\lambda$ and direction $\hat{\omega}$.
For the case of unpolarised non-relativistic radiative transfer, we can write
\begin{equation}
	\phi_{l}(\lambda, \hat{\omega}) = \varphi_l\left(\lambda + \cfrac{v \cdot \hat{\omega}}{c}\lambda_0\right),
\end{equation}
where $v$ is the plasma bulk velocity, and $\lambda_0$ is the rest wavelength of line $l$.
We treat $\varphi$ as a Voigt profile also depending on a spatially- and wavelength-varying damping term $a(\lambda)$, and a spatially-varying Doppler width $\Delta\lambda_D$.
}

\Replaced{The line emission and absorption coefficients can then be rewritten in terms of directionally-independent terms $\eta^*_l$ and $\chi^*_l$}{
Noting that the line profile is the only directionally-dependent component of $\eta_l$ and $\chi_l$, we can write the line emission and absorption components in terms of directionally-independent terms $\eta^*_l$ and $\chi^*_l$
}
\begin{align}
	\eta_l(\lambda, \hat{\omega}) &= \eta^*_l(\lambda) \phi_l(\lambda, \hat{\omega}), \\
	\chi_l(\lambda, \hat{\omega}) &= \chi^*_l(\lambda) \phi_l(\lambda, \hat{\omega}).
\end{align}
\Deleted{where
\begin{align}
	\eta^*_l(\lambda) &= \frac{hc}{4 \pi \lambda} n_j A_{ji}, \\
	\chi^*_l(\lambda) &= \frac{hc}{4 \pi \lambda} \left( n_i B_{ij} - n_j B_{ji} \right),
\end{align}
where $h$ is Planck's constant, $c$ is the speed of light, $n_i$ is the atomic level population of the species in level $i$, $A_{ji}$ is the Einstein spontaneous emission coefficient for the transition $j\rightarrow i$, and $B_{ij}$ and $B_{ji}$ are the Einstein coefficients for absorption and stimulated emission, respectively.
Additionally, to evaluate the line profile $\phi_l$ we require the Voigt damping factor $a(\lambda)$, and the doppler width $\Delta \lambda_D$.
These four parameters are computed from the atomic populations in MIP 0 and then successively averaged along with the velocities throughout the MIP chain.}

\Added{Four MIP chains are then constructed for each spectral line contributing to the wavelength $\lambda$ for the quantities $\eta^*_l(\lambda)$, $\chi^*_l(\lambda)$, $a(\lambda)$, $1 / \Delta\lambda_D$.
The IoD metric selecting the MIP level to use in each block is computed by summing the emissivity and opacity from each line at a Doppler shift of $v_D$ such that $v_D \in \left[-||\vec{v}||, ||\vec{v}||\right]$ for each voxel minimises
\begin{equation}
 f(v) = \left| \lambda_0 \left( 1 - \frac{v}{c} \right) - \lambda \right|,
\end{equation}
i.e. Doppler shifted to the wavelength closest to the line core possible in the current voxel.
This metric combines effects from lines at different velocities, representing a conservative approach to the worst-case gradients in emissivity and opacity.}

\Deleted{The maximum MIP level to sample at each location is then computed using the IoD metric discussed in Section~\ref{Sec:Vlm} taking the wavelength closest to the rest wavelength for each line within the range of possible Doppler shifted wavelengths given the calculation wavelength, and bulk velocity of the mipmapped cell.
This is a conservative approach that ensures regions with high velocities that may meaningfully interact with the line in some directions are not averaged away.}

\Added{\subsubsection{Implementation Details}}

In practice, within the \Dex{} code, the formal solution is solved in batches of wavelengths\Replaced{, and thus a set of active transitions across the current batch is stored, using}{ and the MIPs of wavelength-independent quantities can be shared between wavelengths of a batch for which the same set of lines $\mathcal{L}$ are active. The wavelength-dependent quantities needed for this method can be computed as}
\begin{align}
	\eta^*(\lambda) &= \left( \frac{\lambda_0}{\lambda} \right) \eta^*(\lambda_0), \\
	\chi^*(\lambda) &= \left( \frac{\lambda_0}{\lambda} \right) \chi^*(\lambda_0), \\
	a(\lambda) &= \left( \frac{\lambda}{\lambda_0} \right)^2 a(\lambda_0),
\end{align}
\Added{where $\eta^*(\lambda_0)$, $\chi^*(\lambda_0)$ and $a(\lambda_0)$ are the wavelength-independent quantities (for each line) stored in the mipmaps.}

\Added{As this ``Core and Voigt'' technique requires the evaluation of the line profile for each direction and line of the active set at a wavelength, the implementation of this is critical to the performance of the method, especially if it is to be competitive against the velocity interpolation method of Section~\ref{Sec:VelocityInterp}.}
\Replaced{The method of evaluation of the atomic line profile will depend on implementation requirements, but we highlight here the efficiency of the methods described by \citet{schreier_voigt_2018}.
In particular, we employ their ``HUM1WEI24'' method (region I of \citet{humlicek_optimized_1982} with a 24 term polynomial of \citet{weideman_computation_1994} outside this) which yields good accuracy in a relatively simple form.
This form is particularly suited to GPU calculations, as it has a small footprint (that can be inlined into the assembly) and very limited division operations.
}{The multiple branch structure of the implementation of \citet{humlicek_optimized_1982} maps poorly to vector hardware, and we follow the ``HUM1WEI24'' method of \citet{schreier_voigt_2018}. This method uses the simple rational approximation of region I of \citet{humlicek_optimized_1982}, and a 24 term polynomial due to \citet{weideman_computation_1994} outside this.
The latter maps well to modern floating-point hardware, efficiently exploiting the fused multiply add instruction.}

\section{Performance \& Accuracy Comparison}\label{Sec:PerformanceComparison}

\begin{figure}
	\centering
	\includegraphics[width=\columnwidth]{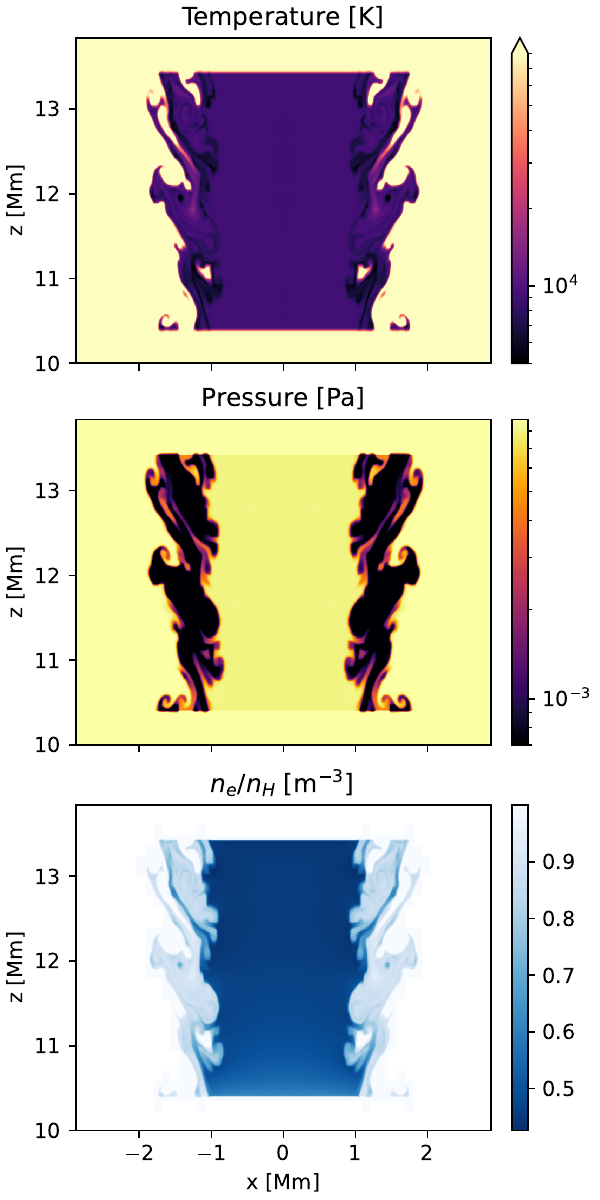}
	\caption{Temperature, pressure, and electron density parameters of the test model. Both temperature and pressure are inputs, but the total hydrogen density $n_H$, and electron density $n_e$ are computed by charge and pressure conservation, thus including the effects of photoionisation.}
	\label{Fig:ModelParams}
\end{figure}

To compare the accuracy and performance of these different methods, we perform a non-LTE calculation using \Dex{}.
\Added{This code implements both AVLM methods discussed in Section~\ref{Sec:Dynamic} inside the radiance cascades framework of \citet{osborne_radiance_2025}.}

\subsection{\Added{Model Configuration}}

This calculation is performed using a snapshot of a 2D model similar to that presented in \citet{snow_observational_2025}\Replaced{, however with an artificial prominence-corona transition region rather than periodic boundaries}{. This model represents a uniform thread of cool and dense solar prominence material undergoing Kelvin-Helmholtz instability in the hot and tenuous corona. To obtain a free-standing non-periodic model, we have applied a sharp temperature increase (and density decrease) on the upper and lower boundaries over a few voxels as a coronal interface.}
The parameters of this model are shown in Figure~\ref{Fig:ModelParams} with a spatial resolution of \SI{7.5}{\kilo\metre} and a grid size of $512\times768$.
The electron density ($n_e$) to total hydrogen density ($n_H$) ratio shown in the bottom panel of Figure~\ref{Fig:ModelParams} is computed by charge and pressure conservation as discussed in \citet{osborne_radiance_2025}, and thus is an output of the model.

As in \citet{osborne_radiance_2025}, this model is computed with charge and pressure conservation, using a 5+1 level model for hydrogen and a 5+1 level model for calcium \textsc{ii} and \textsc{iii}, providing a combined model with 880 wavelengths.
The same solar irradiation boundary condition used in \citet{osborne_radiance_2025} is employed, and the structure is considered at an altitude of \SI{10}{\mega\m} above the solar surface.

\Added{Following \citet{osborne_radiance_2025} we configure \Dex{} with a branching factor of 2 (i.e. all cascades are the same size in memory, angular resolution increases by a factor of 4, and radiance interval end distance from probe centre increases by a factor of 4 with each cascade).
To cover the domain we use 6 cascades, with a cascade 0 probe collocated with every full-resolution voxel and using a radiance interval length of 1.5 voxels with 4 angular samples in the $x-z$ plane.
This is augmented with a 4 sample Gauss-Radau quadrature for different inclinations in $y$ to integrate over the hemisphere (and by symmetry the unit sphere given a homogenous $y$-axis with no velocity components).
Our model thus employs 16 rays on cascade 0, or 4 per octant (albeit with the same angle in the $x-z$ plane).
The formal solver assumes a piecewise constant variation of $\eta$ and $\chi$ within each voxel, and continues to use a long characteristics style solver for each radiance interval, but accelerated by our HDDA traversal algorithm in the cases of mipmapping and sparsity.}

The radiative transfer model is iterated until the maximum relative change of atomic level populations is $<10^{-3}$ \Added{using radiative rates and transition matrices computed at every voxel centre of the underlying model. When using mipmaps, these are computed from scratch from from MIP0, which is in turn calculated from the atomic level populations every iteration.
The atomic level populations are updated using the multilevel accelerated lambda iteration (MALI) technique with same-transition preconditioning \citep{Rybicki1992} and a diagonal approximate operator.
The MIP level is computed using an IoD threshold of 1 for both $\LogEmis{}$ and $\LogOpac{}$, whilst always averaging regions for which $\chi\Delta s < 0.25$.
Additionally, we limit the maximum MIP level sampled by each cascade to $[0, 0, 1, 2, 3, 4]$, thus always employing the maximum resolution for nearby contributions and lower resolution proxies for more distant contributions.
This is primarily due to the upper cascades dominating the calculation time in the formal solution.}

\Added{In our model sparsity is implemented by ignoring blocks for which all voxels have a temperature $>\SI{250}{\kilo\kelvin}$, which represents approximately 50\,\% of the model, for all the regions surrounding the cool thread.
The sparsity map is thus shared between all wavelengths in the model, but the MIP level structure is adapted per wavelength batch (8 consecutive wavelengths).} In all cases the model is run with DexRT v0.5.0 \citep{osborne_goobleydexrt_2025}, on a computer with an NVIDIA RTX 4070 and Intel i7 13700, running the NVIDIA HPC toolkit 24.5 on CUDA 12.4.

\subsubsection{\Added{Velocity Interpolation}}

\Added{The model presented here is solved angularly quadrant by quadrant.
For the velocity interpolation method of Section~\ref{Sec:VelocityInterp}, 21 directional bins are used in each quadrant (leading to an effective 84 directional bins over the full range of velocity projections).
This parameter is easily tuned, and increasing the number of bins will often decrease the error, but at the cost of increased memory consumption, MIP generation time, and cache misses when interpolating the grid, consequently further increasing its runtime.}

\subsubsection{\Added{``Core and Voigt''}}

\Added{The ``Core and Voigt'' technique (Section~\ref{Sec:Cav}) is adopted as the default method in \Dex{}, and has a single tunable parameter: the maximum number of overlapping lines present in a wavelength batch.
This parameter represents the maximum number of lines present in the active set $\mathcal{L}$ in \eqref{Eq:EtaTotDir} and \eqref{Eq:ChiTotDir}.
To enable the loops over these transitions to be unrolled, it is set to a compile-time constant, and for these calculations is set to 4.}

\begin{table*}
\resizebox{\textwidth}{!}{%
\begin{tabular}{lllllllllll}
\hline
Model               & Population   & Max Error (\%) & p99.9 & p99.0 & p98.0 & p95.0 & p67.0 & p50.0 & Time per iter (s) & Num iter \\ \hline
DenseClassic        & -       & 0.0          & 0.0   & 0.0   & 0.0   & 0.0   & 0.0   & 0.0   & 95.0              & 86       \\
DenseCav            & H I (1) & 12.39        & 0.295 & 0.146 & 0.101 & 0.055 & 0.012 & 0.007 & 70.0              & 97       \\
DenseCavMips        & H I (1) & 12.35        & 0.445 & 0.305 & 0.233 & 0.116 & 0.040 & 0.033 & 20.2              & 97       \\
SparseClassic       & H I (1) & 1.20         & 0.031 & 0.019 & 0.016 & 0.010 & 0.004 & 0.002 & 62.9              & 87       \\
SparseCav           & H I (1) & 12.58        & 0.294 & 0.146 & 0.101 & 0.056 & 0.012 & 0.007 & 30.7              & 91       \\
SparseVelInterp     & H I (2) & 12.38        & 0.495 & 0.312 & 0.245 & 0.151 & 0.059 & 0.027 & 22.1              & 82       \\
SparseCavMips       & H I (1) & 12.31        & 0.445 & 0.305 & 0.232 & 0.116 & 0.042 & 0.035 & 11.8              & 101      \\
SparseVelInterpMips & H I (2) & 12.36        & 0.889 & 0.636 & 0.507 & 0.297 & 0.083 & 0.042 & 10.0              & 91       \\ \hline
\end{tabular}%
}
\caption{Error distribution of different methods, showing the population with maximum error (zero indexed per species and ionisation stage), along with time per iteration, and number of iterations for the simulation to converge below a maximum relative change of $10^{-3}$.}
\label{Tab:ModelComparison}
\end{table*}

\subsection{\Added{Results}}

In Table \ref{Tab:ModelComparison} we present the average time per iteration, the number of iterations required for convergence, along with the distribution of relative errors compared to the ground truth model (labelled ``DenseClassic'').
The other techniques are labelled either ``Dense'' or ``Sparse'', depending on whether blocks with temperatures $>\SI{250}{\kilo\kelvin}$ \Replaced{are included in the model (as discussed in Section~\ref{Sec:RayAcceleration})}{are skipped}.
The second component of the technique naming scheme is the method employed for computing the emissivity and opacity:
\begin{itemize}
	\item ``Classic'' is a direct brute force calculation of the emissivity and opacity in every cell for every ray direction.
	\item ``VelocityInterp'' refers to the velocity interpolation method discussed in Section~\ref{Sec:VelocityInterp}.
	\item ``Cav'' refers to the ``Core and Voigt'' method described in Section~\ref{Sec:Cav}.
\end{itemize}
Finally, the technique name is either appended with ``Mips'' if the method used multiple MIP levels or only sampled MIP 0.
The remaining columns show the level population in which the maximum relative error occurs, the magnitude of this error in per cent, and several percentiles of this error distribution.

Whilst in all cases other than the ``Classic'', the peak error is $\sim12\,\%$, this is for a vanishingly small fraction of voxels, as the 99.9th percentile of the error is in all cases below $1\,\%$, and typically below $0.5\,\%$.

As anticipated with its reduced complexity, the velocity interpolation method is somewhat faster, but slightly less accurate than the ``Core and Voigt''.
\Added{However it is only $\sim20\,\%$ faster per iteration whilst doubling the error in level populations.}
\Deleted{In these models we employ 84 directional bins (21 in each quadrant, each of which are solved independently).
Increasing this number of bins can reduce the error, but at the cost of increased memory consumption.
As a result this also increases the number of cache misses when interpolating this grid, and also increases the runtime of this technique.
By default we instead employ the ``Core and Voigt'' method, which has only one tunable parameter: the maximum number of overlapping lines that are allowed (representing $\mathcal{L}$ in the unrolled loops implementing equations \eqref{Eq:EtaCav} and \eqref{Eq:ChiCav}), and the code will not run if this value is set too low.
Herein we set this parameter to 4.
We note that the peak error in this model is always in the first or second excited state of hydrogen, and is likely due to a reordering of terms poorly conditioning a single-precision floating point calculation for an edge case.}

Whilst it is model specific, we find in general that the mipmapping, and the ray acceleration it permits, provide a more significant performance increase than sparsity, other than for extremely sparse models.
This model is 50\,\% sparse with all of the structure in a single central region\Added{, and the use of sparsity does make for a signficant program memory reduction}.
The performance gains found when combining sparsity and mipmapping in this model are slightly lower than we have typically, as this model is less sparse than many.
Using the ``Core and Voigt'' technique with mipmapping (``SparseCavMips''), we obtain an $8\times$ speedup relative the ground truth.
In more sparse models this factor can be significantly increased, often exceeding $10\times$.
In all cases, these timing metrics are a division of the total runtime by the number of iterations, thus including the cost of calculating and configuring the refinement at each wavelength batch.

\section{Discussion}\label{Sec:Discussion}

\begin{figure*}
	\centering
	\includegraphics[width=\textwidth]{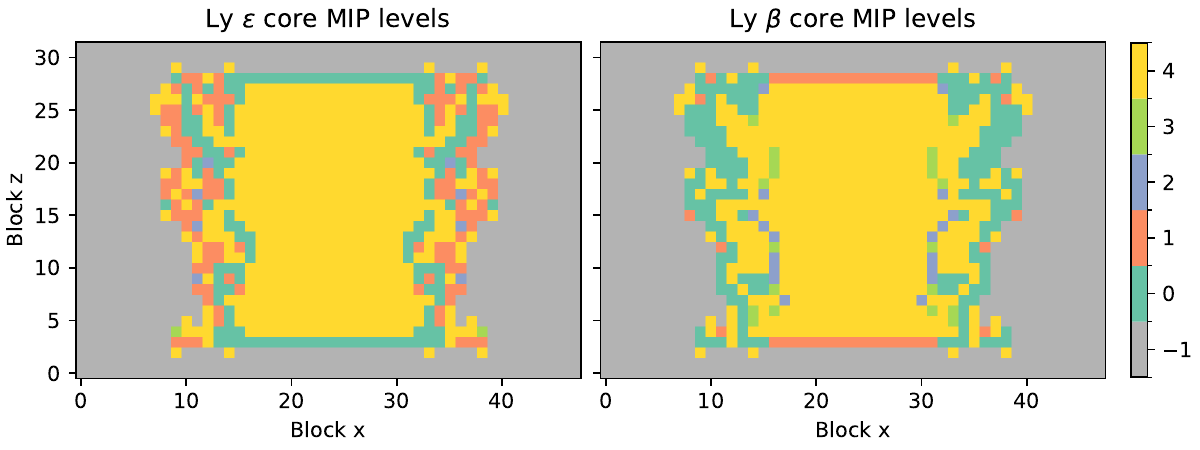}
	\caption{The maximum MIP level for each $16\times16$ block of the model shown in Figure~\ref{Fig:ModelParams}, at the \Added{core }wavelengths \Replaced{of the head of the Lyman continuum (known to be optically thick in prominences) and the}{for the hydrogen Lyman $\epsilon$ and} Lyman $\beta$ spectral lines. A value of $-1$ indicates that the block is empty, and other numbers indicate the maximum MIP level it can be sampled on without adding undue error. It is clear that different resolution is needed in different locations for these two different spectral windows.}
	\label{Fig:MipLevels}
\end{figure*}

\Replaced{This is highlighted}{The per-wavelength adaptivity of the mipmapping scheme is shown} in Figure~\ref{Fig:MipLevels}, where the maximum MIP level per block \Deleted{at the head of the Lyman continuum and }at the core of the \Added{hydrogen Lyman $\epsilon$ and }Lyman $\beta$ spectral line\Replaced{ is}{s are} shown\Added{ for the populations in the converged model using the ``Core and Voigt'' scheme}.
Whilst in both cases the dominant complexity is present in the Kelvin-Helmholtz ``arms'' on the borders of the structure, the resolution requirements are different, with the \Replaced{the Lyman continuum}{Lyman $\epsilon$ line core} able to use a higher MIP level through much of the arms, but requiring a finer solution closer to the core.
On the other hand, Lyman $\beta$ requires the most resolution at the complex interface between the hot and cold material where it typically forms, and thus the emission properties of the plasma will vary rapidly.

The velocity interpolation (Section~\ref{Sec:VelocityInterp}) and ``Core and Voigt'' (Section~\ref{Sec:Cav}) methods shine in \Replaced{massively parallel}{many-core} environments where it is advantageous to calculate rays for multiple directions of the quadrature simultaneously to expose sufficient \Added{parallel} work.
If this is not necessary, the VLM method (Section~\ref{Sec:Vlm}) can be employed to generate the error-limited MIP structure for a grid of emissivity and opacity given a single direction\Added{ (or static atmosphere)}.
This can then be combined with the \Added{adaptive grid and }ray acceleration technique presented in Section~\ref{Sec:RayTraceAcceleration}, to adjust the step size based on the maximum permissible MIP level.

\subsection{\Added{Prefiltering}}

\Added{Mipmapping techniques are a subset of a larger family of methods known as prefiltering.
Prefiltering is the process of modifying input data (typically grid-based) prior to an expensive integration step to reduce the cost of this step (and sometimes reduce error due to aliasing).
Multiple prefiltering techniques exist for treating anisotropic data (i.e. data subject to the traversal problem): for example \citet{crassin_interactive_2011} perform volumetric mipmapping, precomputing the transfer equation (considering only opaque objects) along the $\pm x, y, z$ axes and interpolating between these. \citet{han_frequency_2007} instead prefilter the directional data stored in normal maps by fitting multiple Gaussian lobes.
}

\Deleted{We have also described the concept of prefiltering whereby blocks can be spatially simplified based on traversal direction.}
\Replaced{This concept}{The concept of prefiltering} can be combined with the \Replaced{concept}{use} of radiance intervals and their merging operators from \citet{osborne_radiance_2025} to efficiently combine the average results of stepping through pre-computed blocks of voxels.
\Added{This would effectively extend the work presented here by providing a method that could treat the full RT problem, including iteration as coupled sub-problems with different spatial resolution.}

A recent technique for efficiently handling large quantities of overlapping opaque geometry through voxelisation and prefiltering is presented by \citet{zhou_appearance-preserving_2025}.
\Replaced{It is interesting to note the}{There are interesting} similarities between \Added{this }prefiltering \Added{techhnique }and the hybrid characteristics method \citep{rijkhorst_hybrid_2006}: there are likely gains to be made from employing prefiltering computed from the short characteristics step of hybrid characteristics to improve the performance and coupling of this method when applied to the general case of non-LTE radiative transfer.

\section{\Added{Conclusions}}

In this work we have presented different methods for accelerating the formal solution of the radiative transfer equation by both exploiting sparsity and regions of slow variation in \Replaced{emission and absorption parameters}{emissivity and opacity}.
The methods presented are computed per iteration and wavelength, effectively representing a spatially- and wavelength-adaptive scheme to the structure of the \Replaced{emissive parameters}{emissivity and opacity (directly due to the atomic level populations)}\Deleted{ in a given spectral region} at each iteration.
\Added{These techniques are found to provide an order of magnitude speedup in the formal solution within the radiance cascades based \Dex{} code.}

\Replaced{Furthermore, the}{We note that the} AVLM method as presented here could serve as a key component of an efficient domain discretisation scheme for radiance cascades: lower cascades can be computed locally to each patch, with a full resolution border of ghost cells, whilst higher cascades can be computed on a compacted representation of the mipmap structure, communicating only the data associated maximum level necessary per block.
\Replaced{The}{These} techniques presented \Replaced{here can}{could} also be combined with swept methods such as those presented by \citet{juvela_loc_2020} to provide a scheme where \Replaced{the}{a separate radiative transfer} grid is efficiently adapted to the needs of the \Replaced{emission parameters}{emissivity and opacity}, rather than the structure needed \Replaced{to solve the underlying fluid model}{by the fluid solver}.

Finally, it is important to note that the techniques presented in this work are not exhaustive.
\Replaced{Instead, they demonstrate a method that has regularly provided an order of magnitude computing speedup in our calculations} {We have presented a simple application of prefiltering for the RTE and demonstrated its effectiveness} with the hope of encouraging others to look at potential overlaps between the field of radiative transfer and its neighbours such as computer graphics.

\section*{Acknowledgements}

CMO enjoyed and this work benefited greatly from insightful discussions with Alexander Sannikov and all the regulars of the radiance cascades discord server.
They are also grateful to the Royal Astronomical Society's Norman Lockyer Fellowship, and the University of Glasgow's Lord Kelvin/Adam Smith Leadership Fellowship for financially supporting this work.
\Added{The author also thanks the reviewers for their detailed comments and suggestions for clarification.}

\section*{Data Availability}

The configuration and data produced by these runs is available on Zenodo at \url{https://doi.org/10.5281/zenodo.16759894}.

\section*{Conflicts of Interest}

The author declares no conflict of interest.



\bibliographystyle{rasti}
\bibliography{references}







\bsp	
\label{lastpage}
\end{document}